\documentstyle[12pt,twoside]{article}
{\markboth{\sc Abdelwaheb CHARFI and Salah
HORCHANI}{\sl On the New-Stein Group and the Mass Dynamical Origin}}
\def\C{\hbox { {\rm C}\kern-0.6em\hbox{{\rm I}\ }}}
\def\R{\hbox { {\rm R}\kern-0.9em\hbox{{\rm I}\ }}}
\def\Z{\hbox { {\rm Z}\kern-0.4em\hbox{{\rm Z}\ }}}
\def\N{\hbox {{\rm N}\kern-0.9em\hbox{{\rm I}\ }}}
\def\H{\hbox {{\rm H}\kern-0.9em\hbox{{\rm I}\ }}}
\begin{document}
\noindent{\bf \huge On the New-Stein Group and the Mass  Dynamical
Origin}\vspace{3mm}

\noindent{Abdelwaheb CHARFI}\\ {\small {\it D\'epartement de
Math\'ematiques, Facult\'e des Sciences de Sfax, BP 802, 3018
Sfax, Tunisie}} \vspace{5mm}\\ Salah HORCHANI
\footnote{Corresponding author.{\it E-mail address :
Salah.Horchani@fst.rnu.tn}} \\ {\small {\it D\'epartement de
Math\'ematiques, Facult\'e des Sciences
 de Tunis, Campus Universitaire, 1060 Tunis, Tunisie}}
\vspace{1cm}\\
{\bf Abstract.}
{\small We propose a model of ``Fundamental Symmetries'' leading to a dynamical
origin of mass, based on the concept of  ``Historical Time''  and described by a
relativistic Schr\"odinger-type equation. In this framework, the matter spectrum
is obtained by  excitation, at the level of internal structure, of a fundamental
system composed of two constituents, the relativistic free masses  of
which are  null,  with harmonic interaction
and a possible null-energy vacuum (ground) state}.
\vspace{2mm}\\
{\bf 1991 AMS Subject Classification} : 22E70, 81V25.
\vspace{6mm}\\
{\large {\bf 1. Introduction and Motivations}}
\vspace{2mm}

The combination of the electroweak interactions theory,  developed by
Glashow, Salam and Weinberg [1] and based on the $SU(2)_W \times U(1)_Y$
group, where $SU(2)_W$ (resp  $U(1)_Y)$ is the weak isospin (resp
hypercharge) symmetry, with the strong interactions theory,  generated by the
$SU(3)$-color gauge symmetry, successfully  describes the phenomenology of these
two types of interactions (see, for example,  [2]), to such an extent that this
model has been baptized the ``Standard Model'' of elementary particle
physics. Moreover, this model plays an important role in the discussion of
the primordial universe [3]. However, despite this undeniable success, this
model presents weaknesses. The most important of the latter concerns the
mass observable,  observable which is, from our point of view, the most fundamental,
since it is directly connected to energy, evolution and dynamics. In fact, no
theory related to the Standard Model possesses  a sound ground of mass
generation [4]. All of them rely on Higgs's  mechanism [5] to make the theory
renormalizable and to explain why particles possess a mass. But, this mechanism,
which has been artificially forged for the subsistence of the Standard Model,
 also presents weaknesses. One of the latter is the non-detection, in spite of
considerable efforts for more than thirty years, of the particle necessary for
its survival,  namely the so-called (non-gauge) Higgs boson of spin zero ; this fact
has led, as a matter of fact, to give it an important (lower) mass which would justify
its non-observation at the scale of the presently available energies. Thus, if we
wish to give a sound basis to the origin of mass, we should naturally adopt
other approaches than those relating to the Standard Model.
Everything indicates that this problem
cannot be elucidated without simultaneously establishing an  understanding
of the particles internal structure and a consistent theory of the quantum
relativistic dynamics. It is in this framework that our work should be integrated.

Our approach is based on master ideas formulated by a certain number of
plysicists, inventors of the   elementary particles physics and of modern
cosmology ; these ideas will be briefly exposed below. In  a report published
posthumously by Touschek, Pauli comes back once more to the problem
that ``the fields of all particles are to be constructed from the minimum number
of fields'' [6]. As for Heisenberg, he suggested that it was more correct to
speak of  ``matter spectrum'' rather than  ``elementary particles spectrum''
[7]-[8] (reference [8] concerns the text of Heisenberg's last lecture), and this,
among other things, because the number of particles is not preserved in the
interactions, due to the fact that mass may be  transformed into energy and/or
that a particle-antiparticle pair may be  eventually created. Besides, for Heisenberg, the matter spectrum can only be apprehended if the
sub-adjacent interactions are clarified and the first step must be the attempt
to fromulate mathematically a natural law that defines the dynamics of matter.
In other words, it is the matter dynamics which constitutes the central problem :
``The  dynamics must be taken seriously, and we should not be content with
vaguely defined hypotheses that leave essential points open... The particle
spectrum can be understood only if the underlying dynamics of matter is
known'' [8]. In this context, he suggests to replace the concept of a ``Fundamental
Particle'' by the concept of a ``Fundamental Symmetry''. The fundamental
symmetries define the underlying law which determines the spectrum
of elementary particles. He considers as fundamental symmetries the external
symmetries and the isospin symmetry. As a matter of fact, since its introduction
by Heisenberg [9], the isospin symmetry, charateristic of strong interactions
(and transforming, for example, the proton into neutron), has been
incorporated, on the basis of a Pauli's idea [10], by Heisenberg [11] into his
Nonlinear Spinor Theory of Elementary Particles. It has been extended,
afterwards,  to the weak interactions,  by analogy, into a weak isospin (see,
for example [12]) transforming, for example, a charged lepton into its
neutrino. Heisenberg explicitly excludes the $SU(3)$ symmetry
and its generalizations as fundamental symmetries, because
they may be produced by the dynamics as approximate symmetries. In  fact, the violation of the baryonic number  is a necessary
hypothesis in all unification models of leptons and quarks connected to the
Standard  Model [13]. This violation would explain the asymmetry between
matter and antimatter in the universe [14]. Besides, the baryonic  number
non-conservation is always accompanied  by leptonic   number non-conservation [15]. Finally, Heisenberg
asserts that this decisive change in concepts came about by Dirac's
 discovery of antimatter before concluding : ``... I do not think that we need
any further breakthrough to understand the elementary-or rather non-
elementary-particles. We  must only learn to work with this new and unfortunately
rather abstract concept of the fundamental symmetries...'' [7]. In the discussion
following  the article [7], Dirac adhered to the idea that the concept of  ``elementary
particle'' has no real meaning. Dirac also rejects the legitimacy of the
renormalization procedure as an essential ingredient in the present standard
theories and thinks it is a simple calculation trick, corresponding to  an
incomplete knowledge of the nature of interactions, which ``...does not
conform to the high standard of mathematical beauty that one would expect
for a fundamental physical theory, and leads one to suspect that a
drastic alteration of basic ideas is still needed'' [16]. As for the first rigorous
 approach of the origin of mass, it was born with the inflationary models of
the universe [17]. In these models, the scenario is radically different from
that of the Big Bang  Standard Model until about $10^{-34}$ second  of the
age of the universe. But, beyond $10^{-30}$ second, both scenarios coincide and
all the Big Bang successes are preserved. Let us briefly clarify the {\it raison
d'\^etre}  of these inflationary models. As successful as the Big Bang
cosmology is, it suffers from diverse dilemmas among which we can quote the
initial data and the extreme overproduction of superheavy magnetic
monopoles which occurred early ($t \leq 10^{-34}$ second) in the history of
the universe. It is when trying to solve this type of dilemma that Guth
introduced his idea of inflation in his seminal paper [18]. Let us note that
 there is presently no standard model of inflation, just as, actually, there is
no standard model for physics at these energies (typically $10^{15} GeV$). But
all the inflationary models lead to the fact that all the matter of the universe
burst out from almost nothing : this would be an {\it ex nihilo} creation
as a consequence of a phase transition.  A further
advantage, and not the least, of these inflationary models, is that the evolution
of the universe is nearly independent of the initial conditions which may be
practically arbitrary, because the universe was, prior to the inflationary era,
almost void of matter. This would allow us to avoid the difficulties linked to the
initial singularity of the Big Bang Standard Model. The hypothesis of the creation
of the universe from the absolute nothingness originated from  Jordan's idea
 that the total energy of our universe is null [19], an idea which was afterwards
resumed by Tryon [20] before being integrated within the inflationary theory,
which made it plausible. Tryon suggested that the universe could originally
have been only a spontaneous quantum fluctuation which would have developed
from nothingness. Nothingness in the hypothesis of this  {\it ex nihilo} creation
could designate the  universe at null total energy as Jordan
puts it, or  ``quantum vacuum''. One of our hypotheses is this creation of
the  {\it ex nihilo}  matter to which we give a pure Lie  group framework, and
this from a null-energy ground state associated to a new
relativistic covariant harmonic oscillator formalism (developed through a relativistic  covariant
method unifying the external  \linebreak and the internal spaces which avoids superfluous
 generators)  where the treatment of time as a dynamical variable and the
internal dynamics are expli-\linebreak cit ;  mass (energy)  being created by the internal excitations.

The choice of our formalism  has been motivated by the fact that the
(non relativistic) harmonic oscillator has frequently served as the best laboratory
for theoretical physicits and has served as the first concrete solution to many new
physical theories [21], [22]. Using harmonic oscillators one may construct models
at all levels of complexity : from a single classical oscillator in one dimension
to relativistic quantum fields, passing by the statistical mechanics, the theories
 of specific heat,  superconductivity,  coherent light
etc. Moreover, several works  were carried out concerning the construction of a
non-trivial  relativistic covariant harmonic oscillator wave function for studying,
essentially,  hadronic stuctures and interactions (see, among others, [22] and
[23]). However, several criticisms  can be leveled towards all these attempts ;  the
most important of which bear on the violation of one of the basic  canons of the
relativity theory which stipulates a perfect symmetry between space and time.
 In order to satisfy this fundamental symmetry principle, we shall be compelled to
distinguish, in our treatment, kinematical  time as  observable of relativistic
space-time from dynamical  time, an independant parameter of  this  space-time.
 This dynamical  time is not an observable : It is the Hamiltonian itself which is
the true  unkwown  to determine  and which governs it as a parameter
describing evolution. In this context, the time-evolution constitutes a one
parameter group relativistically compatible with the considered fundamental
 symmetries. We shall determine the  structure of the
various relativistic covariant Hamiltonians
from a very general principle. One of these Hamiltonians will lead  to the new
 relativistic covariant harmonic oscillator formalism  which we referred to above.
Furthermore, the fact that this model permits defining a denumerable infinity
 of fermions (or of bosons) of integer or half-integer isospin (in an irreducible
representation) and that its Hamiltonian presents a symmetry between the
``internal moments'' of its two constituents, which are ``canonically conjugated''
, relates our model to Born's hypotheses concerning his ``Theory
of Reciprocity'' [24], a theory which was approved by Pauli as soon as it
appeared and which Born considers as being ``the unique means to unify the
undulatory mechanics and relativity''. It is worth mentioning here that our
Hamiltonian looks as being a generalization of Born's fundamental invariant.
Furthermore, the possibility that there exist, in this framework,
an infinite number of  different hadrons (and
leptons) does not appear presently unrealistic. But in the framework of the
Standard Model, such a situation is intenable, since it demands an infinite
number of different quarks.

Some years ago  [25], we proposed a model of elementary particles based on
a new concept of internal structure and a relativistically covariant method of
unifying the external and internal structures.  This model led to
 an exact mass
formula for hadrons, compatible with the experimental results. Prior to that
[26], we had introduced a new symmetry group of relativistic quantum
kinematics
(defining the covariance of the moment-energy and position observables)
which was baptized [27] ``Einstein Group'', and which has led, since then,
to diverse applications (for a bibliography concerning this subject, see Chapter
I of [28]). In [28], the non-relativistic equivalent of Einstein Group was called
``Newton Group''.

In this paper, we are going to adopt the concept of internal structure and
 that of  unification, as well  as the construction procedure of the unifying group,
referred to above, by considering the internal structure, reduced to the isospin
as suggested by Heisenbeg, described, as in [25], by the extended
 space Newton  group, and the external structure described by Einstein group.
 Consequently, the  obtained unifying group will be called ``New-Stein Group''.

Thus, in our model, the internal structure is supposed to be, essentially, non-relativistic. As a matter of fact, there is no reason, either theoretical
 or experimental, why the conditions verified by the internal and external
structures should be of the same nature. Furthermore,  the question of the
validity of the usual space-time concept at the scale of elementary particles
 has been put forward very often [29], [30].

By adopting  an evolution principle (unifying dynamics with
 fundamental symmetry  defined by a Lie group $S$), generalizing  the
 one introduced in [31],  and the concept of
``Historical Time'' defined in [32] (which is distinct from the geometrical time
of the relativistic space-time ;  this geometrical  time having, a priori, no relation
with evolution),
 we introduce, in the context of the New-Stein group symmetry, a dynamical
principle of internal evolution which leads us to a relativistic Schr\"odinger-type
equation. It is a model of mass generation which gives a dynamical
explanation of its origin. This is done from a composite fundamental system,
 having two constituents with  a free mass which  may be null and  a harmonic
interaction, the internal excitations of which lead to the creation of the matter
spectrum. Besides, the fact that this model leads to a unitary theory of mass
(energy) creation,  both for the massive and the massless particles (and this as
composite objects),  can lead to a new approach of the unification of the  ultimate
constituents of matter,  in relation with the models of leptons, quarks and gauge
hosons, as composite objects [33], as well as with the supersymmetrical
models [34].

The idea of considering the massless particles as being composite particles
was introduced, from the outset of quantum mechanics,  by Louis de Broglie [35],
in the framework of his ``Photon Undulatory Mechanics'' where he considers
the photon as being a complex particle consisting of two  constituents of $1/2$
 spin. This may be regarded as an early forerunner of the supersymmetrical theories.
This idea had no immediate follow-up , but it has regained intense activity since
the beginning of the  eighties  [36] ; it has led to various models such as that of
the ``Singleton Theory'' [37]. As for the central place reserved by our model to
the massless particles (or constituents), it is motivated, on the one hand, by
the fact that each fundamental interaction of Nature has its privileged family of
massless particles (the photon for the electromagnetic, the neutrinos for the
weak, the gluons for the strong and the graviton for the gravitational) and, on the
other hand, by the fact that, in the inflationary models, the mass generation is
due to a phase transition and that, historically, prior to  the first phase transition,
matter was in its most symmetrical state and almost all elementary particles
were massless [3], [17].
\vspace{3mm}\\
{\large{\bf  2. Definition and Structural Properties of New-Stein Group}}
\vspace{2mm}

Let $E = SL(2, \C) \bullet (\R^4 \times \R^{\,'4})$ the Einstein group [26], inhomogenization
 of $SL(2, \C)$ relatively to its representation $D(1/2,1/2) \oplus D(1/2, 1/2)$,
where the symbols $\bullet$ and $\times$ designate, respectively, the semi-direct
product and the direct product of Lie groups. The Heisenberg group
(resp the Newton group [28]) will be denoted by $H_n$ (resp $F_n$) and the
(universal covering group of the)
extended  space Newton  group by  $F$. We have $F = SU(2) \bullet H_3$,
where the semi-direct product is defined by the representation
$\Delta \oplus \Delta \oplus \varepsilon$ of $SU(2), \Delta$ designating the fundamental representation
and $\varepsilon$ the trivial one. In what follows, Greek indexes run from 1 to 4,
Roman indexes from 1  to 3, summation convention for a repeated index
(in two distinct  positions) is implied and $g_{\alpha \beta}$ is the usual
metric tensor $(g_{ij} = \delta_{ij}, g_{4i} = 0$ and $g_{44} = -1)$. Except when clearly stated, the Lie
groups will be designated by $A, B,\cdots$ and the corresponding Lie algebras
by ${\cal A, \cal B}\cdots {\cal U}({\cal A})$ designates the enveloping algebra
of ${\cal A}$. Let $(M_{\mu\nu}, P_\mu, P'_\mu)$ [resp $(p_i, q_j, I), (I_{ij})]$ the
 canonical basis of ${\cal E}$
(resp ${\cal H}_3, {\cal S} {\cal U}(2))$  and ${\cal G}$ the subalgebra of the tensor product
${\cal U}({\cal E}) \otimes {\cal U}({\cal F})$ generated by :
$$ L_{\mu\nu} = M_{\mu\nu} \otimes I ;  \quad T_\mu = P_\mu \otimes I ; \quad
T'_\mu = P'_\mu \otimes I ;  \quad Q_{j\mu} = P_\mu \otimes q_j ;$$
$$A_{j\mu} = P_\mu \otimes p_j ;  \quad C_{\mu\nu} = P_\mu P_\nu \otimes I ;
 \quad J_{ij} = 1 \otimes I_{ij}.$$

Let ${\cal N}_3$ the subalgebra of ${\cal G}$ generated by
$(A_{j\mu}, Q_{i\nu}, C_{\mu\nu})$. If we denote by $\oplus$ the
direct sum of Lie algebras and by $D(j, j')$ (resp $D(j))$
the irreducible representation of ${\cal SL}(2, \C)$ whose dimension is $(2j+1)(2j'+1)$
(resp of weight $j$ of ${\cal SU}(2))$, then,  we have the following proposition.
\vspace{2mm}\\
{\bf  Proposition 2.1}. ${\cal G}$ {\it is semi-direct sum of}
${\cal S}{\cal L}(2,\C) \oplus {\cal S}{\cal U}(2)$ {\it by the nilpotent ideal}
$\R^4 \oplus \R^{\,'4} \oplus {\cal N}_3$ {\it relatively to the representation} :
$\{D(1/2, 1/2) \otimes D(0)\} \oplus \{D(1/2, 1/2) \otimes D(0)\} \oplus
\{D(1/2, 1/2) \otimes D(1)\} \oplus \{D(1/2, 1/2) \otimes D(1)\} \oplus
\{[D(1, 1) \oplus D(0, 0)] \otimes D(0)\}$.
\vspace{2mm}\\
{\bf Proof} : This is a consequence of  Proposition 2.1 of [25] and of the
 commutation relations of ${\cal G}$, the non-null of which are yielded by :
$$[L_{\mu\nu}, L_{\rho\sigma}] = - g_{\mu\rho}L_{\nu\sigma} -
g_{\nu\sigma} L_{\mu\rho} + g_{\mu\sigma} L_{\nu \rho} + g_{\nu\rho}L_{\mu\sigma} ;
 [L_{\mu\nu}, X_\rho] = g_{\nu\rho}X_\mu - g_{\mu\rho}X_\nu ;$$
$$[A_{i\mu}, Q_{j\nu}] = \delta_{ij} C_{\mu\nu} ;
[L_{\mu\nu}, C_{\rho\sigma}] = - g_{\mu\rho} C_{\nu\sigma} - g_{\mu\sigma}
C_{\nu\rho} + g_{\nu\sigma} C_{\mu\rho} + g_{\nu\rho}C_{\mu\sigma} ;$$
$$[J_{ij}, J_{kl}] = - \delta_{ik} J_{jl} - \delta_{jl}J_{ik} + \delta_{il} J_{jk} +
\delta_{jk} J_{il} ; [J_{ij}, Y_k] = \delta_{jk} Y_i - \delta_{ik} Y_j ;$$
where ${X}_\rho$ (resp $Y_k$) designates $T_\rho, T'_\rho, A_{i\rho}$
 or $Q_{i\rho}$ for i fixed (resp $A_{k \rho}$,  $Q_{k\rho}$ for $\rho$ fixed).
\vspace{2mm}

Let $\R^{10}_c$ (resp$\R^{12}_a, \R^{12}_q$) the group generated by
$(C_{\mu\nu})$ (resp $(A_{j\mu}), (Q_{j\mu}))$. Now,
 we shall denote by $t ({\rm resp}\   t', c, a, q, \wedge, R)$ the generic element of \linebreak
$\R^4 ({\rm resp} \R^{\,'4}, \R^{10}_c,  \R^{12}_a, \R^{12}_q, SL(2, \C), SU(2))$,
by $\theta^{\mu \nu}$ the function equal to $1/2$ if $\mu = \nu$ and 1 if not and
by $g =(t, t', c, a, q, \wedge, R)$  the generic element of the connected and simply
connected group $G$ whose Lie algebra is ${\cal G}$.In what follows, we note
$\wedge$ (resp$R, S(\wedge))$ instead of $D(1/2, 1/2) \wedge ({\rm resp}
 \;D(1) R, [D(1, 1) \oplus D(0, 0)] (\wedge) \otimes D(0)R).$ .
\vspace{2mm}\\
{\bf Proposition 2.2}. {\it The group law of}  $G$ {\it is given by}
$g_1g_2 = (t_1 + \wedge_1 t_2, t'_1 + \wedge_1 t'_2, c_1 + S(\wedge_1) c_2 +
\beta(a_1, \wedge_1 \otimes R_1q_2), a_1 + \wedge_1 \otimes R_1a_2, q_1 +
\wedge_1 \otimes  R_1 q_2, \wedge_1\wedge_2, R_1R_2)$,
{\it where} $\beta$ {\it is defined by}  $\beta^{\mu\nu} (a_1, q_2) = \theta^{\mu\nu}
\delta_{ij} (a^{i\mu}_1 q^{j\nu}_2  + a^{i\nu}_1 q^{j\mu}_2).$
\vspace{2mm}\\
{\bf Proof} :  Results form Proposition 2.1. and from the group law of $N_3$
[25]  whose Lie algebra is ${\cal N}_3.$
\vspace{2mm}\\
{\bf Definition 2.1.} {\it Group}  $G$  {\it is called the New-Stein Group}.
\vspace{3mm}\\
{\large{\bf 3. On a Class of Irreducible Unitary Representations
 of the New-Stein Group}}
\vspace{2mm}

Proposition 2.2. shows that $G$ admits the decomposition $H\bullet K$ where
$H = \R^4 \times \R^{\,'4} \times \R^{10}_c \times \R^{12}_a$
and $K = \R^{12}_q \bullet (SL(2, \C) \times SU(2))$.
Consequently, the most natural method of determining its
strongly continuous  irreducible unitary representaions (IUR) is the method of
induced representations [38] (in stages),  provided that it turns out to have
the required properties. Let $\hat{H}$ be the dual of $H$, then
$\hat{H} = \R^4_p \times \R^4_\xi \times \R^{10}_D \times
\R^{12}_b$. As in [25], in order to determine the action
of $K$ on  $\hat{H}$ , we write $D$ and $b$ as  a matrix. Thus, the element
$(p, \xi, D, b)$ of $\hat{H}$ will be represented by
 $(p, \xi, DG, B)$, where
$D = (D^{\mu\nu})$, with $D^{\mu\nu} = D^{\nu\mu}$,
$G = (g_{\alpha\beta})$ and  $B = (b^{i\mu})_{1\leq i\leq 3; 1\leq \mu\leq 4}
 \in M(4, 3, \R)$.
Thus, the action of $ k = (0, 0, 0, 0, q, \wedge, R) \in K$
 on $\hat{h} =(p, \xi, DG, B) \in \hat{H}$ is defined by :
$$k(\hat{h}) = (\wedge p, \wedge \xi, \wedge DG^{-1} \wedge, \wedge BR^{-1},
\wedge DG^{-1} \wedge Q) = (p', \xi', D'G, B') ; \eqno{(1)}$$
where $Q$ denotes the matrix $(q^{i\mu})_{1\leq i \leq 3, 1 \leq \mu \leq 4}
 \in M(4, 3, \R)$.

We note that the action of $K$ on $\R^4 \times \R^{\,'4}$ is reduced to
the action of $SL(2, \C)$.
\vspace{2mm}\\
{\bf Remark 3.1.} In view of the physical interpretations we wish to draw from
our model, we shall confine ourselves in this article to the determination
of IUR of $G$ for which : $(g_{\nu\mu} p^\mu p^\nu = 0, p^4 > 0)$ and
 $(g_{\nu\mu} \xi^\mu \xi^\nu = - m^2_0, m_0 > 0)$. Thus, the IUR
searched for are those which are associated to the orbits  contained in
$\Omega_+^0 \times \Omega_+^{m^2_0}.$
\vspace{2mm}\\
{\it 3.1. Determination of the stabilizer}
\vspace{2mm}

It results from what preceded that the orbits looked for are written :
$$(S_\lambda \times \{\lambda\}, \Omega^{m_0^2}_+),  \mbox{ with  }
m_0 > 0 \mbox{ and  } \lambda > 0 ;\eqno{(2)}$$
where $S_\lambda$
is the sphere having a radius  $\lambda$ in the hyperplan $p^4 = \lambda$.

They are generated by $(\vec{\eta}, \lambda, (0, 0, 0, m_0))$.
 For all $\vec{\eta} \in S_\lambda$, the
stabilizer  of point $(\displaystyle\mathop{p}^o,
\displaystyle\mathop{\xi}^o) = ((0, 0, \lambda, \lambda), (0,0,0, m_0))$
is the subgroup $\R^{12}_q \bullet (U(1) \times SU(2))$
 the action of which on $\R^{10}_D$ is  given, according to (1),
by $(q, \wedge, R) (DG) = \wedge DG \wedge^{-1}$. Let the
$\displaystyle{\mathop{D}^o}G$ matrix, all the elements of which are null except
$(\displaystyle{\mathop{D}^o}G)^{44} = - \alpha < 0$.
 Then, we have a point-orbit, since
we have $\wedge \displaystyle{\mathop{D}^o} G\wedge^{-1} =
\displaystyle{\mathop{D}^o} G, \forall\; \wedge \in SU(2)$.
The orbit deduced from the action of $K$
on $\R^4_p \times \R^4_\xi \times \R^{10}_D$ is thus :
$$(S_\lambda \times \{\lambda\}, \Omega^{m^2_0}_+, \{\displaystyle{\mathop
{D}^o} G\}), \quad \lambda > 0 ; \eqno{(3)}$$
it is generated by the point
 $(\displaystyle{\mathop{p}^o}, \displaystyle{\mathop{\xi}^o},
\displaystyle{\mathop{D}^o}G)$ stabilized by the subgroup
$\R^{12}_q \bullet (U(1) \times SU(2))$. Let us study now the action of this subgroup on
$\R^{12}_b$. According to (1) we have :  $\forall\; (q, \wedge, R) \in
\R^{12}_q \bullet ((U(1) \times SU(2)), (q, \wedge, R) (\displaystyle{\mathop{p}^o},
\displaystyle{\mathop{\xi}^o}, \displaystyle{\mathop{D}^o} G, B) =\linebreak
(\displaystyle{\mathop{p}^o}, \displaystyle{\mathop{\xi}^o},
\displaystyle{\mathop{D}^o}G, B')$ with
 $B' = \wedge BR^{-1} + \wedge \displaystyle{\mathop{D}^o}
G^{-1} \wedge Q$. Let us write
$B = \left(\matrix{
{\bf b}\cr
{^t}\vec{\beta}} \right)$,  with ${\bf b} = (b^{i\mu})_{i\leq 3, \mu \leq 3}$,
where $\vec{\beta}$ is the column vector $(b^{i4})_{i\leq 3}$
 and ${^t}\vec{\beta}$
 the transposed of $\vec{\beta}$ ; hence, knowing that $\wedge \in U(1)
\subset SU(2)$, then
of the form $\left( \matrix{ &(\wedge(\varphi)) &0\cr
&0 &1}\right)$,  where $\wedge(\varphi)$ is the canonical matrix associated
 to the rotation of angle $\varphi$  around the third space axis,  we have :
$B' = \left( \matrix{
{\bf b'}\cr
{^t}\vec{\beta}'}\right) =
\left( \matrix{
\wedge (\varphi) {\bf b}\cr
{^t}\vec{\beta}}\right) R^{-1} + \displaystyle{\mathop{D}^o} G \left( \matrix{
{\bf q}\cr
{^t}\vec{q}}\right)$, where we have written,
like $B, Q = \left( \matrix{ {\bf q}\cr
{^t}\vec{q}}\right)$ ; hence
  $B' = \left( \matrix{
\wedge(\varphi) {\bf b} R^{-1}\cr
{^t}(R {\vec \beta} - \alpha \vec{q})} \right)$.
The orbit searched for is generated by $\displaystyle{\mathop{B}^o}
= \left( \matrix{
0\cr
{^t}\vec{\beta}}\right)$ . It is then defined  by
$\{B' = \left( \matrix{
0\cr
{^t}(R\vec{\beta} - \alpha\vec{q})}\right) / R \in SU(2),  \vec{q} \in \R^3\}$,
 that is $\left\{\left( \matrix{
0\cr
{^t}\vec{\beta}}\right) / \vec{\beta} \in I\!\!R^3 \right\}.$
 The orbit $\Omega$ to  consider is generated by
$\hat{h}_0 = (\displaystyle{\mathop{p}^o}, \displaystyle{\mathop{\xi}^o},
\displaystyle{\mathop{D}^o}G, 0)$ and
 finally written :
$$\Omega = (S_\lambda \times \{\lambda\}, \Omega^{m^2_0}_+ ,
\{\displaystyle{\mathop{D}^o}G\},
\left\{ \left( \matrix{
0\cr
{^t}\vec{\beta}}\right) / \vec{\beta} \in  \R^3\right\}). \eqno{(4)}$$

An element $(q, \wedge, R) \in \R^{12}_q \bullet (U(1) \times SU(2))$,
 where $q$ is identified with  matrix
$Q = \left( \matrix{ {\bf q}\cr
{^t}\vec{q}}\right)$, stabilizes  $\displaystyle{\mathop{B}^o} = 0$ if and only
if we have $\displaystyle{\mathop{D}^o}GQ = 0$. But
$\displaystyle{\mathop{D}^o}GQ = \left( \matrix{ 0\cr
-\alpha {^t}\vec{q}}\right)$
and $\alpha \neq 0$ ; hence the stabilizer of $\hat{h}_0$ is :
$$K_0 = \R^9_q \bullet (U(1) \times SU(2)) ; \eqno{(5)}$$
where $\R^9_q = \{(q^{i\mu}) \in \R^{12}_q$ so as $q^{i4} = 0\}$.
\vspace{2mm}\\
{\it 3.2. Construction of  section}
 $\Gamma_{\hat h}$ {\it and calculation of}
$\Gamma_{\hat h}^{-1}k_0 \Gamma_{k_0^{-1}(\hat{h})}$ {\it  for}  $\hat{h} \in
\Omega$ {\it and}  $k_0 \in K$
\vspace{2mm}

 Each point of  orbit $\Omega$  is written :
 $\hat{h} = (q, \wedge, R) \hat{h}_0 = (\wedge \displaystyle{\mathop{p}^o}
, \wedge \displaystyle{\mathop{\xi}^o},\displaystyle \wedge \displaystyle{\mathop{D}^o}
G^{-1} \wedge,\linebreak \wedge \displaystyle{\mathop{D}^o}G  \wedge^{-1}Q).$
If we pose
$Y = \wedge^{-1} Q = \left(\matrix{
\tilde{y}\cr
{^t}\vec{y}}\right)$, then
$\displaystyle{\mathop{D}^o}GY = \left( \matrix{
0\cr
-\alpha {^t}\vec{y}}\right)$
and   ${^t}\vec{y}$ is
invariant by the subgroup  $K_0$ (that is to say ${^t}\vec{y}$ is the same if
we replace $(q, \wedge, R)$
by $(q, \wedge, R)k_0$,  with $k_0 \in K_0)$. If we
denote for $\xi \in \Omega_+^{m^2_0}$ and $\vec{\eta} \in S_\lambda$,
  $A_\xi$ and $R_{\vec \eta}$ the habitual
sections defined in the canonical formalism, then each point of  orbit
$\Omega$ is written  :
$\hat{h} = (A_\xi R_{\vec \eta} \displaystyle{\mathop{p}^o}, A_\xi
 \displaystyle{\mathop{\xi}^o}, A_\xi \displaystyle{\mathop{D}^o}G A^{-1}_\xi,\linebreak
A_\xi
\left( \matrix{
0\cr
- \alpha{^t}\vec{y}}\right))$ .
As $\Omega$ is a variety of
dimension 8 (isomorphic with  $K/K_0)$, a section associated to our orbit
is given by $\Gamma_{\hat h} = (A_\xi
\left( \matrix{ 0\cr
{^t}\vec{y}}\right), A_\xi R_{\vec\eta},1)$,  where 1 is the neutral element of $SU(2)$ . It
is to be noticed that  orbit $\Omega$ is parametered by
$(\xi, \vec{\eta}, \vec{y}) \in \Omega^{m^2_0}_+ \times S_\lambda \times \R^3$.
Hence : $\Gamma^{-1}_{\hat h} = (- \left(\matrix{ 0\cr {^t}{\vec y}}\right),
R^{-1}_{\vec \eta} A^{-1}_\xi, 1)$.

Let $k_0 = (Q_0, \wedge_0, R_0) \in K$ , with matrix writing  of the
component  of
$k_0$ belonging to
$\R^{12}_q$,  then $k^{-1}_0 = (- \wedge_0^{-1} Q_0R_0, \wedge^{-1}_0,
R^{-1}_0)$ and we  \linebreak obtain : $k^{-1}_0(\hat{h}) = (\wedge^{-1}_0 A_\xi
R_{\vec \eta}\displaystyle{\mathop{p}^o},
 \wedge^{-1}_0 A_\xi \displaystyle
{\mathop{\xi}^o}, \wedge^{-1}_0 A_\xi \displaystyle{\mathop{D}^o} G A^{-1}_\xi
\wedge_0, \wedge^{-1}_0 A_\xi \left( \matrix{ 0\cr
- \alpha {^t}\vec{y}}\right) R_0\linebreak
 - (\wedge^{-1}_0 A_\xi \displaystyle{\mathop{D}^o}
G A^{-1}_\xi \wedge_0) (\wedge^{-1}_0 Q_0 R_0))$.
But,  $A^{-1}_\xi \wedge_0 A_{\wedge^{-1}_0 \xi} \in SU(2)$
(since it stabilizes $\displaystyle{\mathop{\xi}^o}$), then it stabilizes
$\displaystyle{\mathop{D}^o}G$ ; thus
$k^{-1}_0 \hat{(h)} = (A_{\wedge^{-1}_0 \xi} R_{\vec \eta'}
\displaystyle{\mathop{p}^o}$
($\vec{\eta'}$ to determine), $A_{\wedge_0^{-1}\xi} \displaystyle{\mathop{\xi}^o}$,
$A_{\wedge_0^{-1}\xi}\displaystyle{\mathop{D}^o}G A^{-1}_{\wedge_0^{-1}\xi}$,
$A_{\wedge_0^{-1}\xi}(\left( \matrix{0\cr
- \alpha{^t}\vec{y}}\right) - \displaystyle{\mathop{D}^o}G A^{-1}_\xi Q_0) R_0 = d).$
Let  $A^{-1}_\xi Q_0 = A^{-1}_\xi (q^1_0, q^2_0, q^3_0) = \left( \matrix{ T\cr
{^t}\vec{\tau}}\right)$,  where  $\vec{\tau} \in \R^3.  {^t}\vec{\tau} = \{(A^{-1}_\xi
q^1_0)^4, (A^{-1}_\xi q^2_0)^4,\linebreak (A^{-1}_\xi q^3_0)^4\}$ ;
which yields
$d = A_{\wedge_0^{-1}\xi} \left( \matrix{ 0\cr
- \alpha {^t}(\vec{\tau} - \vec{y})}\right)$.

Then,  we have obtained :
$k^{-1}_0 \hat{(h)} = (A_{\wedge_0^{-1} \xi} R_{\vec \eta'}\displaystyle{\mathop{p}^o},
A_{\wedge_0^{-1}\xi}\displaystyle{\mathop {\xi}^o},
A_{\wedge_0\xi}
\displaystyle{\mathop{D}^o}G A^{-1}_{\wedge^{-1}_0 \xi}, \linebreak A_{\wedge^{-1}_0\xi}
\left( \matrix{ 0\cr
- \alpha{^t}(\vec{y} - \vec{\tau})}\right) R_0)$ ;
the last component may be written \\
$A_{\wedge_0^{-1}\xi} \left( \matrix{
0\cr
- \alpha{^t}(R^{-1}_0 (\vec{y} - \vec{\tau}))}\right)$
with $\vec{\eta'} =  D(1) (A^{-1}_{\wedge_0^{-1}\xi}
\wedge^{-1}_0 A_\xi) \vec{\eta}$.
Consequently,  $\Gamma_{k_0^{-1}\hat{(h)}} = (A_{\wedge_0^{-1}\xi}
\left( \matrix{
0\cr
{^t}R^{-1}_0(\vec{y} - \vec{\tau})}\right), A_{\wedge_0^{-1}\xi}
R_{\vec \eta}, 1)$.
In conclusion, taking into account the fact that $A^{-1}_\xi \wedge_0 A_{\wedge_0^{-1}\xi}
\vec{\eta} \in SU(2))$, we obtain :
$$ \Gamma^{-1}_{\hat h} k_0 \Gamma_{k_0^{-1}(\hat{h})} = (R^{-1}_{\vec \eta}
\left( \matrix{
T\cr
0}\right),
R^{-1}_{\vec \eta} A^{-1}_\xi  \wedge_0 A_{\wedge_0^{-1}\xi}
R_{\vec \eta}, R_0).\eqno{(6)}$$
{\bf Remaks.3.2.}  (a) Orbit  $\Omega$ being homeomorphic with
 $\Omega^{m_0^2}_+ \times S_\lambda \times \R^3$,
the action of $k = (Q, \wedge, R) \in K$ on $\hat{h} = (\xi, \vec{\eta}, \vec{y})$
 is defined by $k^{-1}(\hat{h}) = (\wedge^{-1} \xi, D(1) (A^{-1}_{\wedge^{-1}\xi}
\wedge^{-1} A_\xi) \vec{\eta}, R^{-1}(\vec{y} - \vec{\tau}))$,
so as $A^{-1}_\xi Q = \left( \matrix{ T\cr
{^t}\vec{\tau}}\right) \in M(4, 3, \R)$.

(b) The decomposition $G = H \bullet K$ is also written
 $G = (\R^4 \times G_1) \bullet K$,  where
$G_1 \bullet K$ is the group considered in [25] Section 6, which is a regular semi-direct
product, and the action of $K$ on $\R^4$ is reduced to the standard action
of $SL(2, \C)$ ; it yields the habitual orbits of   the Poincar\'e group.
Then, the product $H \bullet K$ is regular.

(c) If a group $G$ is a semi-direct product of two closed unimodular subgroups
$H$ and $K$, $H$ being normal in $G$, it follows from [39],  Chapter II Paragraph 7,
that $G$ is unimodular if,  for every function $f \in {\cal L}^1_\mu(H)$,  we have
$\int_H f(k h k^{-1}) d\mu(h) = \int_H f(h) d\mu(h)$, where $d\mu$ is an invariant
 Haar measure on $H$. It follows, on applying this property in stages, that all
the semi-direct products encountered are unimodular and that their invariant
measures are obtained by simply considering the product of the measures  of
their factors.

\noindent{{\it 3.3. IUR of the New-Stein Group associated to}
$\Omega$} \vspace{2mm}

Let $L$ an IUR of the stabilizer $K_0 = \R^9_q \bullet (U(1) \times SU(2))$
defined on the Hilbert space ${\cal H}_L$ and
trivial on $\R^9_q$, then $L$ is of the form :
$L = D^{(s)} \otimes D(j)$, where $D^{(s)}$ denotes
a unidimensional
representation of $U(1)$ on $\C$ with $s \in \{0, \pm \frac{1}{2}, \pm 1, \cdots \}$.
It follows that $L$ is defined on
$\C \otimes \C^{2j+1} = \C^{2j+1}$ by : for all
$(Q, \wedge, R) \in K_0$ and $v \in \C^{2j+1}$,
 $L(Q, \wedge, R) v =  D^{(s)}(\wedge) \otimes D(j)(R)v$. Then, we have the following
Theorem.
\vspace{3mm}\\
{\bf Theorem 3.1.} {\it The IUR of} $G$ {\it induced by the  IUR of} $K_0 =
\R^9_q \bullet (U(1) \times SU(2))$, {\it associated to
the trivial orbit of} $U(1) \times SU(2)$ {\it in} $\R^9_q$, {\it are given  by} :
$\{U(t, t', c, a, q, \wedge, R)F\} (\xi, \vec{\eta}, \vec{y}) = \exp i\{ \langle p, t\rangle
+ \langle\xi, t'\rangle + \langle b, a\rangle_1 +  \langle D, c\rangle_2\} \times$\\
$D^{(s)}(R^{-1}_{\vec \eta} A^{-1}_\xi \wedge A_{\wedge^{-1}\xi} R_{\vec \eta'})
\otimes D(j) (R) (F(\wedge^{-1} \xi, \vec{\eta'}, R^{-1} (\vec{y} - \vec{\tau}))),$\\
{\it where} : $p = A_\xi R_{\vec \eta}\displaystyle{\mathop{p}^o}$,
$b = B = A_\xi \left( \matrix{
0 \cr
- \alpha{^t}\vec{y}}\right), DG = A_\xi \displaystyle{\mathop{D}^o} GA^{-1}_\xi,\\
F \in {\cal H}  = {\cal L}^2_\mu (\Omega_+^{m^2_0} \times S_\lambda \times \R^3,
\C^{2j+1})$,  $d\mu = dw (\xi) d\sigma(\vec{\eta})dz$,
{\it with} $dw$(resp $d\sigma$) {\it being the invariant
measure by} $SL(2, \C)$ ({\it resp} $SU(2))$ {\it concentrated
on the  hyperboloid} $\Omega^{m^2_0}_+$ ({\it resp the sphere}
$S_\lambda$),  $\vec{\eta'} = D(1) (A^{-1}_{\wedge^{-1}\xi} \wedge^{-1}A_\xi)
\vec{\eta}$, $\langle x_1, y_1 \rangle = g_{\mu\nu} x_1^\mu y_1^\nu, \langle b, a\rangle_1
= \delta_{ij} g_{\mu\nu} b^{i\mu} a^{j\nu}$ {\it and}
$\langle D, c\rangle_2 = \sum_{\mu\leq \nu} D_{\mu\nu} c^{\mu \nu}$.
\vspace{3mm}\\
{\bf Remarks 3.3}. (a) If $\vec{q}^\nu$ is the column  vector $(q^{i\nu})_{i\leq 3}$,
$\vec{\tau} = - \frac{1}{m_0} g_{\mu\nu} \xi^\mu \vec{q}^\nu$
 and consequently $R^{-1}(\vec{y} - \vec{\tau}) = R^{-1}(\vec{y} +
\frac{1}{m_0} g_{\mu\nu} \xi^\mu \vec{q}^\nu)$.

(b) If $R^{-1}_{\vec \eta}  A^{-1}_\xi \wedge A_{\wedge^{-1}\xi}R_{\vec \eta}$
is the inverse image of $\wedge(\varphi)$, by the universal covering
homomorphism, whose eigenvalues are $e^{\pm i \varphi/2}$,
then  for all
$ g \in G, F \in {\cal H}$ and $2s \in \Z,  \{U(t, t', c,
a, q, \wedge, R)F\} (\xi, \vec{\eta}, \vec{y}) =  \exp i \{\langle p, t\rangle +
\langle\xi, t'\rangle + \langle b,a\rangle_1 + \langle D, c\rangle_2 + s\varphi\} \times
 D(j) RF(\wedge^{-1} \xi, \vec{\eta}, R^{-1} (\vec{y} + \frac{1}{m_0} g_{\mu\nu}
\xi^\mu \vec{q}^\nu)$ ;
 or, by writing $\vec{z} = - m_0\vec{y}$, $(U(g) F) (\xi, \vec{\eta}, \vec{z}) = \exp
 i \{\langle p, t\rangle + \langle \xi, t'\rangle + \langle b, a\rangle_1 + \langle D, c\rangle_2
 + s\varphi\} D(j) RF(\wedge^{-1} \xi, \vec{\eta}, R^{-1} (\vec{z} - g_{\mu\nu}
\xi^\mu \vec{q}^\nu))$.
\vspace{2mm}

Differentiating this representation, we obtain for infinitesimal generators
(defined on the space of its ${\cal C}^\infty$-vectors) :
\vspace{2mm}\\
\begin{eqnarray*}
T'_\mu &=& \xi_\mu ;\quad  T_\mu = p_\mu
\mbox{ such that }  p = A_\xi R_{\vec \eta}p_0; \quad
C_{\mu\nu} = \frac{\alpha}{m^2_0} \xi_\mu \xi_\nu ;\\
A_{j\mu}& =& \frac{\alpha}{m^2_0} z^j \xi_\mu  ;\quad   Q_{j\mu} =
i \frac{\partial}{\partial z^j} \xi_\mu ;\\
 J_{23} &=& S^1_j + i(\vec{z} \bigwedge
\frac{\partial}{\partial \vec{z}})^1 ;\;\; J_{31} = S^2_j + i(\vec{z} \bigwedge
\frac{\partial}{\partial \vec{z}})^2 ;\;\; J_{12} = S^3_j + i(\vec{z} \bigwedge
\frac{\partial}{\partial\vec{z}})^3 ;\\
 L_{23} &=& \frac{s\eta^1}{k + \eta^3} +
i\{(\vec{\xi} \bigwedge \frac{\partial}{\partial \vec{\xi}})^1 + (\vec{\eta} \bigwedge
\frac{\partial}{\partial\vec{\eta}})^1\} ;\\
L_{31} &=& \frac{s\eta^2}{k + \eta^3} +
i\{(\vec{\xi} \bigwedge \frac{\partial}{\partial \vec{\xi}})^2 + (\vec{\eta} \bigwedge
\frac{\partial}{\partial\vec{\eta}})^2\} ;\\
L_{12} &=& s + i\{(\vec{\xi} \bigwedge \frac{\partial}{\partial \vec{\xi}})^3 + (\vec{\eta}
\bigwedge \frac{\partial}{\partial \vec{\eta}})^3\} ;\\
L_{14} &=& \frac{s[\eta^2 \xi^3 - \xi^2(k+\eta^3)]}
{(k + \eta^3)(m_0 + \xi^4)} + i\{\frac{\xi^3 \eta^3 + \xi^2 \eta^2}
{\xi^4 + m_0} \frac{\partial}{\partial\eta^1} - \frac{\eta^1\xi^2}{\xi^4+m_0}
\frac{\partial}{\partial \eta^2}\\
&-& \frac{\eta^1\xi^3}{\xi^4 + m_0} \frac{\partial}{\partial \eta^3} + (\xi^4
\frac{\partial}{\partial \xi^1} + \xi^1 \frac{\partial}{\partial \xi^4})\} ;\\
L_{24} &=& \frac{s[-\eta^1 \xi^3 + \xi^1(k + \eta^3)]}{(k+\eta^3)(m_0 + \xi^4)}
 + \{ - \frac{\xi^1 \eta^2 + \xi^1 \eta^2}{\xi^4 + m_0} \frac{\partial}{\partial \eta^1}
+ \frac{\eta^1 + \xi^1 + \xi^3 \eta^3}{\xi^4 + m_0} \frac{\partial}{\partial \eta^2} \\
&-& \frac{\eta^2\xi^3}{\xi^4 + m_0} \frac{\partial}{\partial \eta^3} + (\xi^4
\frac{\partial}{\partial \xi^2} + \xi^2 \frac{\partial}{\partial \xi^4})\} ;\\
L_{34} &=& \frac{s[\eta^1 \xi^2 - \xi^1\eta^2]}{(k + \eta^3)(m_0 + \xi^4)} +
i\{- \frac{\eta^3}{\xi^4 + m_0} (\xi^1 \frac{\partial}{\partial \eta^1} + \xi^2
\frac{\partial}{\partial \eta^2}) + \frac{\eta^1 \xi^1 + \eta^2 \xi^2}
{\xi^4 + m_0} \frac{\partial}{\partial \eta^3}\\
&+& (\xi^4 \frac{\partial}{\partial \xi^3} + \xi^3  \frac{\partial}{\partial \xi^4})\} ;
\end{eqnarray*}
where $(S^l_j)_{l\leq 3}$ denote the infinitesimal generators of the
representation $D(j)$ of $SU(2)$, $\vec{\xi}$ the vector $(\xi^i)_{1\leq i \leq 3}$
 and $\bigwedge$ the vectorial product.
\vspace{4mm}\\
\noindent{\bf 4. Cohomology and Deformations of the New-Stein Lie Algebra }
\vspace{2mm}

The deformations theory [40] appears in a very natural way in symmetry
 contexts, mainly  to discover all the possible symmetries that are connected,
in some sense, to the given one. It also permits a better understanding of the
transitions between the fundamental levels of physics (non-relativistic,
relativistic, classical, quantum...). For instance,  the  passage from classical Newton
mechanics to classical relativistic mechanics can be interpreted as a
deformation of the Galilei group into the Poincar\'e group (which, in turn, may be
 deformed into the de Sitter group).
The deformations of
a Lie algebra can be searched for systematically by computing its cohomology
groups. This cohomological calculation will also allow us, in Paragraph 5, to
determine what we shall call the  {\it relativistic invariant extensions}  of $\R$  by  the
New-Stein Lie algebra. These extensions are at the basis, in Paragraph 7, of the
 unification of the non-relativistic and relativistic dynamics by making the latter
a Newtonian-type dynamics.

The results of this paragraph, together with their demonstration, are similar
to those of [41], {\it Mutatis Mutandis}. The non-explicited notations are those
of [42].

Let  $Z({\cal SL}(2, \C) \oplus {\cal SU}(2))$
(resp $Z({\cal G}))$ the centralizer of
${\cal SL}(2, \C) \oplus {\cal SU}(2)$ in ${\cal G}$ (resp the center of
${\cal G}$).
\vspace{2mm}\\
{\bf Lemma 4.1}. {\it The three vector spaces}
 $H^0 ({\cal G, G}), Z({\cal SL}(2, \C) \oplus {\cal SU}(2))$ {\it and}
$Z({\cal G})$ {\it are unidimensional and
generated by} $C = g^{\mu\nu} C_{\mu\nu}$.
\vspace{2mm}\\
{\bf Lemma 4.2.} {\it For}  $i \in \{1, 2\}$,  {\it the} ${\cal G}$-{\it modules}
$H^i({\cal G, G})$  {\it and} $H^i(\R^4 \oplus \R^{\,'4} \oplus {\cal N}_3,
{\cal G})^{\cal G}$ {\it are isomorphic}.
\vspace{2mm}\\
{\bf Lemma 4.3.} (a) {\it Any element} $f \in Z^1 (\R^4 \oplus \R^{\,'4} \oplus {\cal N}_3,
{\cal G})^{\cal G}$
 {\it may be parametred by} :
$f(T_\mu) = \alpha T_\mu , f(T'_\mu) = \alpha'T'_\mu, f(A_{i\mu}) = \beta
A_{i\mu} + \gamma Q_{i\mu}, f(C_{\mu \nu}) = (\beta + \gamma')C_{\mu\nu},
f(Q_{i\mu}) = \beta' A_{i\mu} + \gamma' Q_{i\mu}$,  {\it where}
$\alpha, \alpha', \beta, \beta', \gamma$ and  $\gamma'$ {\it are real numbers}.

(b) $H^1(\R^4 \oplus \R^{\,'4} \oplus {\cal N}_3, {\cal G})^{\cal G}$ {\it is trivial}.
\vspace{2mm}\\
{\bf Theorem 4.1.} $H^1({\cal G, G})$ {\it is a} ${\cal G}$-{\it module of dimension 6}.
\vspace{2mm}\\
{\bf Lemma 4.4.} (a) {\it Any element} $g \in B^2(\R^4 \oplus \R^{\,'4} \oplus {\cal N}_3,
{\cal G})^{\cal G}$ {\it is defined}\linebreak {\it by} : $g(X, Y) = 0$ {\it for all } $X, Y \in \{L_{\mu\nu},
T_\mu, T'_\mu, C_{\mu\nu}, J_{ij}\}$ {\it and}
 $g(A_{i\mu}, Q_{j\nu}) = \delta_{ij}(\alpha C_{\mu\nu} + \beta g_{\mu \nu}
g^{\rho \sigma} C_{\rho\sigma})$, {\it where} $\alpha, \beta$ {\it are
real numbers}.

(b) $Z^2(\R^4 \oplus \R^{\,'4} \oplus {\cal N}_3,  {\cal G})^{\cal G} = B^2(\R^4 \oplus
 \R^{\,'4} \oplus {\cal N}_3, {\cal G})^{\cal G}$.
\vspace{2mm}\\
{\bf Theorem 4.2.} $H^2 ({\cal G, G}) = \{0\}.$
\vspace{2mm}\\
{\bf Corollary 4.1.} ${\cal G}$ {\it is a rigid Lie algebra}.
\vspace{4mm}\\
{\bf 5. New-Stein Group Relativistic Invariant Extensions }
\vspace{2mm}

As  first basic hypothesis of evolution in the framework of fundamental symmetry
defined by a Lie group $S$,  we suppose that evolution is described by a one-parameter Lie
group $\R$ , a subgroup of a Lie group $\tilde{S}$ (unifying fundamental
symmetry and dynamics) so that $S$ is a normal subgroup of $\tilde{S}$
and the quotient group $\tilde{S}/S$ is $\R$ . Which defines $\tilde{S}$ as being an
extension of $\R $   by $S$ . In order to  search for the possible $\tilde{S}$ groups, we
are going to reason at the level of Lie algebras and to determine, afterwards,
the corresponding Lie groups.

An extension of $\R$    by ${\cal S}$ is defined by an exact sequence :
$0 \rightarrow {\cal S} \stackrel {\lambda}{\rightarrow}\tilde{\cal {S}}
\stackrel {\mu} \rightarrow \R \rightarrow 0$
so as $\lambda ({\cal S}) = {\it Ker} \mu$. Such an extension is inessential [43],  since
$\R$ is one-dimensional, and it is defined by a linear mapping,  associating
 to the generator $K$ of $\R$ the derivation $\Phi$ of ${\cal S}$, given by :
$\forall\; x \in {\cal S}, \Phi(x) = [K, x]$.

 As second basic hypothesis, we suppose that $K$ is invariant by any purely external or purely
internal symmetry. We call  such an extension a {\it relativistic invariant extension}.

If we apply these results to fundamental symmetry defined by the New-Stein
group $G$, we have $\Phi \in Z^1(\R^4 \oplus \R^{\,'4} \oplus {\cal N}_3, {\cal G})^{\cal G}$ and,
in Lemma 4.3.,   $\alpha = \alpha' = 0$. Then, we
obtain the following Proposition.
\vspace{2mm}\\
{\bf  Proposition 5.1.} {\it Any relativistic invariant extension} $\tilde{{\cal G}}$ {\it is defined}
{\it by the commutation relations} :
 $[K, A_{i\rho}] = \beta A_{i\rho} + \gamma Q_{i\rho} ; [K, Q_{i\rho}] = \beta'
A_{i\rho} + \gamma' Q_{i\rho}$ ; \linebreak $[K, C_{\mu\nu}]
 = (\beta + \gamma') C_{\mu\nu}$ ;
{\it the other non-null commutation relations
being those of} ${\cal G}$.
\vspace{2mm}\\
{\bf Remark 5.1.} In order to classify such extensions, we consider the
matrix $L = \left( \matrix{
\beta &\beta'\cr \gamma &\gamma'}\right)$ associated to the restriction of {\it ad} $K$ at the subspace generated
by $A_{i\rho}, Q_{i\rho}$ ($i, \rho$ fixed). We notice that if $L$
 is equivalent to $L'$,  then the associated
extensions are equivalent ; consequently, we treat $L$ under  reduced form
of Jordan. Let $P$ the characteristic polynomial of $L, P = X^2 - X {\it tr} L +
{\it det} L$ ; if {\it det} $L \neq 0$,
the transformation $K' = K : \sqrt{|det L|}$ does not change the brackets of ${\cal G}$ ;
besides, if $L'$ is associated to $K'$, we have det $L' =$\linebreak $\pm 1$  ; hence, we have
six cases defined by
$({\it det} L \in \{0, 1, -1)$ {\it and} $({\it tr} L = 0$ {\it or} ${\it tr} L \neq 0))$,
 which yield 9 Jordan reduced forms for $L$
(following the values of {\it tr}  $L$), leading to 9 non-equivalent  extensions of
$\R\;$   by ${\cal G}$, analogous to those defined in [31]. This leads to  the following
Proposition.
\vspace{2mm}\\
{\bf Proposition 5.2.} {\it Any relativistic invariant extension} ${\tilde{\cal G}}$
{\it is equivalent to one of the  Lie algebras defined by the following Lie brackets} :

(1)
$[K, A_{i\rho}] = 0 ;\;\;\;\qquad  [K, Q_{i\rho}] = 0 ; \;\;\;\qquad  [K, C_{\mu\nu}] = 0$ ;

(2)
$[K, A_{i\rho}] = A_{i\rho} ; \qquad  [K, Q_{i\rho}] = - Q_{i\rho} ; \quad  [K, C_{\mu\nu}] = 0$ ;

(3)
$[K, A_{i\rho}] = \zeta^2 A_{i\rho} ; \quad  [K, Q_{i\rho}] = - \zeta^{-2} Q_{i\rho} ;\  \quad
[K, C_{\mu\nu}] = (\zeta^2 - \zeta^{-2})  C_{\mu\nu}$ ;\\
 { \it where }  $\zeta^2 \notin  \{0,1\} ;$

(4)
$[K, A_{i\rho}] =  \zeta^2 A_{i\rho} ; \!\!\!\!\!\qquad  [K, Q_{i\rho}] = \zeta^{-2} Q_{i\rho} ;
 \qquad  [K, C_{\mu\nu}] = (\zeta^2 + \zeta^{-2})  C_{\mu\nu}$ ;\\
 { \it where }  $\zeta \neq  0$ ;

(5)
$[K, A_{i\rho}] = \zeta^2  A_{i\rho} ; \quad  [K, Q_{i\rho}] = 0 ; \quad
 [K, C_{\mu\nu}] = \zeta^2 C_{\mu\nu} ; { \it where }\; \zeta \neq 0$ ;

(6)
$[K, A_{i\rho}] = - Q_{i\rho} ; \quad [K, Q_{i\rho}] = 0 ; \quad  [K, C_{\mu\nu}] = 0$ ;

(7)
$[K, A_{i\rho}] = - Q_{i\rho} ; \quad  [K, Q_{i\rho}] = A_{i\rho} ;  [K, C_{\mu\nu}] = 0$ ;

(8)
$[K, A_{i\rho}] = \cos \varphi A_{i\rho} - \sin \varphi Q_{i\rho} ; \quad  [K, Q_{i\rho}]
= Q_{i\rho}  ; \quad [K, C_{\mu\nu}] = 2C_{\mu\nu}$ ;

(9)
$[K, A_{i\rho}] = \cos \varphi A_{i\rho} - \sin \varphi Q_{i\rho} ; \quad [K, Q_{i\rho}] =
\sin \varphi A_{i\rho} + \cos \varphi Q_{i\rho}$ ;\\
 $[K, C_{\mu\nu}] = 2 \cos
\varphi C_{\mu\nu} ; { \it where }\; \varphi  \neq (k + 1/2)\pi$.

{\it The other Lie brackets being those of ${\cal G}$}.
\vspace{2mm}\\
{\bf Remark 5.2.} In order to determine the Lie groups corresponding to these
Lie algebras, it is sufficient to  apply the Campbell-Hausdorff formula [44].
In Paragraph 7, we shall study, in detail, the fundamental symmetry
associated to the relativistic invariant  extension defined by case (7) the group of which
is given by the following Proposition.
\vspace{2mm}\\
{\bf Proposition 5.3}. {\it The group law of} $\tilde{G}$ {\it (whose Lie algebra is defined by
case (7) of Proposition} 5.2.){\it  is given by :
 for all} $g_i = (k_i, t_i, t'_i, c_i, a_i, q_i, \wedge_i, R_i) \in \tilde{G} (i = 1, 2)$,
 {\it then} $g_1g_2  = (k_1 + k_2, t_1 + \wedge_1 t_2, t'_1 + \wedge_1 t'_2, c_1 + S(\wedge_1) c_2 +
(\sin^2 k_2) \times \linebreak \beta(a_1, q_1) - \frac{1}{4} (\sin 2k_2) [\beta(a_1, a_1) - \beta
(q_1, q_1)] - \beta(q_1 \cos k_2 + a_1 \sin k_2, \wedge_1 \otimes R_1a_2), \linebreak
a_1 \cos k_2 - q_1 \sin k_2 + \wedge_1 \otimes R_1 a_2, q_1 \cos k_2 + a_1 \sin k_2 + \wedge_1
\otimes R_1 q_2, \wedge_1 \wedge_2, R_1R_2).$
\vspace{4mm}\\
{\bf 6. Remarkable IUR of an Extension of $\R\;$   by the New-Stein Group}
\vspace{2mm}

Let $\tilde{G}$ the extension of $\R$ by $G$ defined in Proposition 5.3. ; then
$G$ is a normal  (non-abelian) subgroup  of ${\tilde G}$ and  to determine
the IUR  of $\tilde{G}$ we apply the induction method [38]. Let us fix its notations :
 Let $\chi \in \hat{G}$, then $k \in \R$  acts on $\chi$ by  :
$\forall g \in G, (k \chi)(g)= \chi(kg k^{-1})$. Let $O({\cal X})$ the orbit of
 $\chi$ and $K_\chi$ the stabilizer
of $\chi$. If $(\rho, {\cal H}_\rho)$ is a IUR of $K_\chi$ and ${\cal H}_\chi$
the support
of $\chi$, we obtain a representation $\pi$ of $K_\chi \bullet G$, defined
on ${\cal H}_\rho \otimes {\cal H}_\chi$, by writing
$\pi(k, g) = \rho (k) \otimes W(k) \chi(g)$, where $W(k)$
is the isomorphism of  ${\cal H}_\chi$ realizing the
equivalence of the representations $\chi$ and $k \chi$. From this
 IUR of $K_\chi \bullet G$,
we obtain, by induction, an IUR of $\tilde{G}$. Besides, all the IUR
of $\tilde{G}$ are of this form.
\vspace{2mm}\\
{\bf Proposition 6.1.} {\it The IUR of} $\tilde{G}$,
{\it  induced by an IUR}  $(U, {\cal H})$  {\it of}
$G$ {\it of the type obtained in Paragraph 3 and by the character}
$k \rightarrow e^{ik \ell/2}$ {\it of} $\R$ , {\it is given by}
$$\tilde{U} (k, g) F  = \exp i \frac{k}{2} (\ell - \frac{m^2_0}{\alpha} \Delta +
\frac{\alpha}{m^2_0}  z^{(2)}) U(g)F, $$
{\it where} $(k, g) \in \tilde{G} (= \R \bullet G)$,   $F \in {\cal H}$,
$\Delta = \sum_j \frac{\partial^2}{(\partial z^j)^2}$ {\it and} $z^{(2)} = \sum_j
(z^j)^2$.
\vspace{2mm}\\
{\bf Proof} : Knowing that $k \in \R\;$   commutes with $t, t', c, \wedge$ and $R$, it is
sufficient to consider $(kU)(a)$ and $(kU)(q)$, where $a \in \R^{12}_a$
and $q \in \R^{12}_q$.   Theorem 3.1.  gives :
$U(kak^{-1}) = \exp i\{\cos k a^{j\mu} \frac{\alpha}{m^2_0} z_j \xi_\mu +
\frac{\sin 2k}{4} \frac{\alpha}{m^2_0} \sum_{\mu \leq \nu} \beta(a,a)^{\mu\nu}
\xi_\mu \xi_\mu\} \times \exp i\{- \sin k a^{j\mu} [i \frac{\partial}{\partial z^j}
\xi_\mu]\}$ and
$U(kq k^{-1}) = \exp i \{\sin k q^{j\mu} \frac{\alpha}{m^2_0} z_j \xi_\mu -$ \linebreak
$\frac{\sin 2k}{4} \frac{\alpha}{m^2_0} \sum_{\mu \leq \nu} \beta(q, q)^{\mu\nu}
\xi_\mu \xi_\nu\}
\times \exp i\{ \cos kq^{j\mu} [i \frac{\partial}{\partial z^j}\xi_\mu]\}$ ;
by writing
 $U(a) =$ \linebreak $\exp ia^{j\mu} \frac{\alpha}{m^2_0} [z_j \xi_\mu]$ and
$U(q) = \exp iq^{j\mu} [i \frac{\partial}{\partial z^j} \xi_\mu]$,
we obtain
$U(kak^{-1}) =$ \linebreak $ W(k) U(a) W(k^{-1})$
and  $ U(kqk^{-1})  = W(k) U(q) W(k^{-1})$,
where $W(k)$ is the unitary operator  $W(k) = \exp \frac{ik m^2_0}{2\alpha}
(- \Delta + \frac{\alpha^2}{m^4_0}z^{(2)})$.
Finally $U(kgk^{-1}) =$\linebreak $ W(k) U(g) W(k^{-1}),\; \forall g \in G$,
 and the orbit of $U$ is $\{U\}$, whose stabilizer is $K_U = \R$ .
Consequently,  to the character $k \rightarrow e^{ik\ell/2}$ of $\R\;$  is associated
the IUR  $\;\tilde{U}$
of $\tilde{G}$ defined on $\C \otimes {\cal H}$ by :
 $\tilde{U}(k, g) = \exp \frac{ilk}{2} W(k) U(g)$.
\vspace{4mm}\\
{\bf 7. Physical Interpretation}
\vspace{2mm}

In the same way as the non-relativistic classical mechanics, the standard
quantum mechanics privileges time with regard to space,  which is
contrary to the very spirit of the Einstein  relativity theory. In fact, in quantum
mechanics, time is a parameter, whereas the spatial position observables are dynamical
variables. Similary, in non-relativistic classical mechanics, it is possible to define
the  spatial  position of a particle, whereas it is  not possible to say that a particle has a
well defined time. Moreover, the Minkowski space-time seems to favor the
statical conception of time with regard to its dynamical conception. Thus, if we
 want to re-establish the space-time symmetry in relativistic quantum mechanics,
we must distinguish the geometrical time of the space-time, which is an observable,
associated to a clock, defining the state of the system (time of the event in the
laboratory referential), from dynamical time, an independant parameter
of space-time, which is not an observable, but the function of which is to describe the
evolution of the system. This distinction will permit the unification of the
relativistic and non-relativistic dynamics by turning both of them into a Newtonian
 dynamics : in both cases, the evolution time is managed by a dynamical
concept which makes it  ``Universal Time'' passing ``uniformly and inexorably''
as Newton imagined. Such an additional parameter was introduced by
various authors, but, at the beginning [45], simply, as a mathematical
convenience,  without  physical interpretation. It is only afterwards  that
this evolution parameter was taken into consideration, out of necessity, in order
 to conciliate the ideas of Einstein and those of Newton and to build up a
Hamiltonian-type relativistic dynamics, and this by admitting   the existence of a
space-time in conformity with the geometry defined by the Poincar\'e group, but
occupying, vis-\`a-vis evolution, a position analogous to that of the three-dimensional
space in Newton theory. This parameter has been baptized ``
Historical Time'' [32], and this is not the proper time of
any particular system.

It is in this framework that we shall interpret symmetry in accordance with the
New-Stein group, an interpretation which will permit to give ``Historical Time''
and the dynamics ensuing from it a pure theoretical group basis.

First of all, the generators of ${\tilde{\cal G}}$, other than those of the Poincar\'e
algebra, are relativistic covariant since they are invariant under translations
and transform, under the action of the Lorentz group, like a Lorentz tensor. In
order to define the dynamics associated to the fundamental symmetry sub-jacent
to the New-Stein group, we adopt the following Postulate.
\vspace{2mm}

{\bf Postulate} : {\it To any relativistic invariant extension} $\tilde{G}$,
{\it of the New-Stein
 group} $G$, {\it corresponds a dynamics, generated by the infinitesimal generator}
$K$ {\it of} $\tilde{G}$, {\it the evolution parameter of which is identified  with} \linebreak
{\it ``Historical Time''} $\tau$ .
\vspace{2mm}

At this stage, two important implications of this Postulate deserve mentioning :
the first is that the specification of the dynamics may not be attempted, even
in principle,  before the  symmetry group is decided upon ; as for the second
implication, it stipulates  that symmetry entirely determines the dynamics structure,
 that is to say the concept which governs the system modifications  during time.
According to Weinberg [46], such a connection between symmetry and dynamics
was encountered, for the first time, in the ``Theory of Strings''. In general, we
introduce dynamics in a model by adding an interaction term.

Generator  $K$, consequently, generates evolution in historical time and, being
an invariant  with regard  to anything which is purely external or internal symmetry,
it is preserved during evolution. Therefore, we shall interpret it as being the
total mass (energy) of the system. By its definition, it also appears as being a
relativistic Hamiltonian. It is convenient to note, in this context, that, in general,
the Hamiltonian invariance is slightly stronger than the motion
equation
invariance ; for example, for an isolated system, the Galilei  group does leave
the motion  equation invariant, but alters the Hamiltonian.

In what follows, we are going to study, in detail, the fundamental symmetry
associated to the relativistic invariant extension $\tilde{G}$ defined by Proposition 5.3. and
in the context of its IUR defined by Proposition 6.1. . The privilege of this
fundamental symmetry essentially lies in the fact that it is going to allow us to
generalize the standard formalism of the harmonic oscillator [21]
as well as  Bohr's
fundamental invariant, associated to his ``Theory of  Reciprocity'' [24], in the
relativistic framework with presence of an  internal dynamical  structure.

Taking into account our interpretation, the mass observable of our system is
represented by $d\tilde{U}(K)$, where $d\tilde{U}$ is the differential
representation of the $\tilde{U}$ representation defined on the space of its
$C^\infty$-vectors. We, consequently, obtain :
$$d\tilde{U}(K) = \frac{1}{2} (- \frac{m^2_0}{\alpha} \Delta + \frac{\alpha}
{m^2_0} z^{(2)} + \ell).$$

The operator $d\tilde{U}(K)$ coincides with the representative of element $B$
of the enveloping algebra ${\cal U}\tilde{({\cal G})}$ defined by
$$B = (\frac{1}{2\alpha} M^2_N + \frac{\ell}{4}) + (\frac{1}{2\alpha} M^2_A +
\frac{\ell}{4}), $$
where :  $\qquad M^2_N = - g^{\mu\nu} \delta^{ij} Q_{i\mu} Q_{j\nu},
M^2_A = - g^{\mu\nu} \delta^{ij} A_{i\mu} A_{j\nu}.$

As for the representatives of these two last elements of ${\cal U}(\tilde{\cal G})$
in the representation
$d\tilde{U}$,  they are given by :
$$d\tilde{U} (M^2_N) = - m^2_0 \Delta, d\tilde{U} (M^2_A) = \frac{\alpha^2}{m^2_0} z^{(2)}.$$

The mass observable $d\tilde{U}(K)$ admits a discrete  spectrum leading to the
exact mass formula :
 $$m^2_n = n + \frac{3}{2} + \frac{\ell}{2},$$
 where $n$ is a non-negative integer and
$\ell$ a real number ; whereas the operators $d\tilde{U}(M^2_N)$ and
$d\tilde{U}(M^2_A)$ admit continuous spectra.

This leads us to the following interpretation based, partially, on the ideas developed
in Paragraph 5 of [25]. We interpret the $Q_{i\mu}$ and the $A_{i\mu}$ (and the
$C_{\mu\nu}$ which depend
on them algebraically) as describing the internal dynamics of a composite
system with two constituents  in interaction, conjugated from each other, which we shall call
``Nazzamion'' $N$ and ``Antinazzamion'' $A$ [47],  in the following way :
the Nazzamion (resp Antinazzamion) energy-momentum is  given by
 an ``external'' part $T_\mu$ and an ``internal'' part $Q_{i\mu}$ (resp $A_{i\mu})$.
Thus, the relativistic free mass of its constituents is null, since $d\tilde{U}
(T_\mu T^u) = 0$, and, consequently,  $d\tilde{U}(K)$ effectively represents the
mass (energy) of this composite system. This does not contradict the relativistic
quantum dynamics where the mass of a composite system (for instance, the
mass of the resonance state) can be larger than the sum of the masses  of
the constituents  ; the excess mass can be
visualized as the positive kinetic energy of the constituents-particles in the
potential well. In this framework, the internal degrees of freedom are thus
assumed to contribute to the creation of mass via a term which is (for each
space-time direction) the Hamiltonian of the (three-dimensional) harmonic
oscillator. This term can be viewed as the sum of the self-interactions $M^2_N$
and $M^2_A$ of the Nazzamion and the Antinazzamion ; each of these
self-interaction terms has a continuous spectrum in the representation
$d\tilde{U}$, but the coupling of both  leads to
our composite system  which we shall call ``Nazzamium''.
 The  Nazzamium dynamics  is defined by  the Hamiltonian $d\tilde{U}(K)$
which  has  a discrete  spectrum of the harmonic oscillator-type, capable, by
its nature, of describing the matter spectrum of fundamental symmetry
managed by the New-Stein group. In view of this, we should have a proper
relativistic covariant formalism for the harmonic oscillator mass spectrum
where the basic constituents can be of vanishing mass. As for the isospin
content of this model, it is absolutely analogous with  that of the harmonic
oscillator model developed in paragraph 6.A of [25]. In particular, the
eigenfunctions of $d\tilde{U}(K)$ will be classified by a principal quantum number
connected  with the internal excitation level and a secondary quantum number
associated with the weight of a IUR of the factor $SU(2)$ of the Newton group
(the isospin symmetry). Thus, we arrive to a dynamical  explanation of mass
(energy) origin : it is created by excitation of the Nazzamium internal
structure. An important consequence of this  analysis is that the generation
of mass is associated with the generation of the internal symmetry and that
its origin appears as a bound state effect or an interaction effect linked to
that symmetry. In this context, we must point out that the presence of
oscillations in our model relates it to the Theory of Strings (see, for instance,
[49]).
In fact, in many versions of this theory, the strings are closed into loops, and
it is not these loops that represent particles but the various ways in which
the loops can oscillate. The energies of these oscillations, expressed as
mass by the Einstein equivalence, are the mass of the particle we know.

In a way, this interpretation seems to be a generalization of the standard
description of a  system of $n$ quantum particles possessing only properties
that have a classical analog. As a matter of fact, there exist, for such a system,
$n$ pairs of canonically conjugate variables
$(P_i, Q_i)_{1\leq i \leq n}$, representing
the generators of Heisenberg algebra ${\cal H}_n$, so that the Hamiltonian (and  also
every other physical observable) is a function of it. In our model, ${\cal N}_3$,
which characterizes the internal dynamics, would generalize the
Heisenberg algebra, and  element $B$, associated with $d\tilde{U}(K)$
in ${\cal U}(\tilde{\cal G})$, would generalize the fundamental
invariant of Born [24]. Finally, this interpretation has also the advantage that
none of the generators of ${\tilde{\cal G}}$ is overabundant, contrarily to the various models
related to the Standard Model.
\vspace{2mm}

The New-Stein group symmetry presents four important advantages by
comparison with that developed in [25] :

(a)
It permits, not only, to determine the matter spectrum, but, it also fixes its
 diverse dynamics which are classified by the extensions of the dynamic
 group $\R\;$  by the fundamental symmetry group $G$.

(b) Thanks to the concept of ``Historical Time'' $\tau$, evolution can be described,
rigorously, by a relativistic Schr\"odinger-type equation
 $$i \partial_\tau \Psi_\tau = d\tilde{U}(K) \Psi_\tau,$$
analogous to the standard way to describe the dynamics of a non-relativistic
particle where the change of  states  is given by unitary transformations.
Such unitary transformations are supposed to form a one-parameter group
representation. According to Stone's theorem such an evolution is completely
defined by a self-adjoint operator leading to the Schr\"odinger equation.

(c)
 The energy of the quantum vacuum (fundamental state of the Nazzamium)
may be chosen equal to zero : the only condition to satisfy is to have
$\ell = - 3$ in the mass formula. Consequently,
the ground state is completely devoid of dynamics, but not of kinematics.

(d)
The mass term can be achieved from the very dynamics even when the
constituents are massless.
\vspace{2mm}

Consequently,  the New-Stein group symmetry leads to a unified description
of massless and massive  particles : the passage of the former to the latter
is carried out by excitations at the level of the Nazzamium internal structure.
This might be the basis of a new approach to the unification of all the fundamental
interactions of Nature. In this context, we may suppose that, like in the model
of the {\it ex nihilo}  creation of matter associated  with the inflationary
theory of the primordial universe, the generation of mass in our model is also
a consequence of a phase transition. Thus, there exists a phase structure
between massive and massless particles  : quantum vacuum  goes into a series of phase transitions ; the first of which,
historically speaking (that which is, as  a matter of fact,  located in the most
speculative era of contemporaneous cosmology [3] and during  which the
 three types of strong, weak and electromagnetic interactions are unified),
corresponds to the creation of mass. The estimate of the critical temperature
$T_c$ of this phase transition has  been confirmed by the work of Weinberg
 [50] ; it is approximately $10^{27}K$ and corresponds to a typical energy of
$10^{15}$ {\it GeV} located at the instant $10^{-34}s$ after the Big Bang [51].
At temperature $T > T_c$, mass will disappear and matter is in a highly
 symmetrical state. This evidently suggests that in the primitive universe,  at
very high temperature, there were only massless particles, an epoch which we
may call the {\it massless era}. Later,  the {\it mass era} began at $T = T_c$ due to a
phase transition during which the massive particles appear as a bound  state
of Nazzamium. As for the physical reality of the Nazzamium constituents, they
may possibly be only mathematical fictions, having nothing to do with the
notion of particle as we commonly conceive it, but leading, effectively, to the
fundamental symmetries, as expressed by Heisenberg.
\vspace{4mm}\\
{\bf 8. Discussion and Outlook}
\vspace{2mm}

We end this work by  briefly  making a few remarks and suggestions concerning
 its possible continuations :

(1)
Since the New-Stein group is not a group of  transformations  admitting a passive
interpretation in space-time, in order to study the consequences of the symmetry it
induces, it has not been necessary to determine its unitary or anti-unitary
projective representations, which can be  obtained from  the study of its central
extensions by $\R$ . However, these extensions may lead  us to possible new
symmetry groups which are not ``very far'' from the New-Stein group.
The space of
equivalence classes  of the central extensions of the New-Stein Lie
algebra by $\R$ is of dimension eleven. The most promising of the correspondent
 groups is   group $G_\beta$ associated with the local exponent $\beta$ of the
New-Stein group defined by $\beta(g_1, g_2) = \langle t_1, \wedge_1  t'_2\rangle$,  for
any two elements
$g_i(t_i, t'_i, c_i, a_i, q_i, \wedge_i, R_i), i \leq 2$, of the New-Stein group.
Group $G_\beta$ contains the extended Einstein group and leads, like it [26], to
generalized Heisenberg relations and to a new definition of the relativistic
spin.

(2)
The deformations of the extended Newton groups are analogous to those of the
extended Einstein group [41], {\it Mutatis Mutandis}. Thus, taking into account
the rigidity of the New-Stein group (consequence of Corollary 4.1.) and of the
non-rigidity of the extended Einstein and Newton groups, the replacement,
in this framework, of one of these two groups by one of its deformed groups
would lead to a fundamental symmetry which is  (separately) ``neighboring'' (from the
external or  internal structure view point)  the one studied in our work.
Besides, if we replace the extended space Newton group by that of a higher
dimension (by substituting {\it Spin}$(n)$ for $SU(2)$ and $H_n$ for $H_3$, with
$n > 3)$, the corresponding New-Stein group is still rigid.

(3)
In what was exposed above, the Nazzamium Hamiltonian $d\tilde{U}(K)$
coincides with the representative of element $B$ of the enveloping algebra
${\cal U}(\tilde{{\cal G}})$. Now,  $B$ possesses the property of being a
symmetrical homogeneous polynomial of the second degree in the conjugate
canonical variables $(A_{i\nu})$ and $(Q_{i \nu})$ which describe the internal
dynamics. This, naturally,  suggests the  study of other evolution kinds generated
by such polynomials. Thus, for instance, if we suppose that, at least in a
first approximation, the ``internal'' energy-momentum observables  of the Nazzamion and
the Antinazzamion add linearly, we are led to consider the composite system
the evolution of which is generated by
 $- g^{\mu\nu} \delta^{ij}(Q_{i\mu} + A_{i\mu}) (Q_{j\nu} + A_{j\nu})$ .

There remains, of course, the study of the other dynamics associated with
the other  New-Stein group  relativistic invariant  extensions  of Proposition 5.2.

(4)
It can be intuitively judged that an oscillator is not sufficient to describe a
system as extended object  and, similarly to molecular physics which combines
the oscillator and the rotator to obtain a vibration-rotation energy, it would
be necessary to add a term which would correspond to a quantum relativistic
rotator. For that purpose, the central extension $G_\beta$ of the New-Stein
group can be an ad hoc framework, since it is possible to consider
$S_{\mu \nu} S^{\mu\nu}$ as the  term describing the rotation energy, where
$S^{\mu\nu}$ is the relativistic spin tensor defined in [26] and which has been
constructed by analogy with the Galilean mechanics.

(5)
In another perspective, the fact that the formalism of the creation and annihilation
 operators explicity includes the mass creation concept from energy, it would
be interesting to define (working from internal canonical variables $(A_{i\mu})$
and $(Q_{i\mu}))$ operators $X^\pm_{i\mu}$ of the  creation and
annihilation type and adapt our model to a field theory (with an infinite
number of particles).

(6)
The notion of the dimensionality of space-time is a fact which lies at the very
foundations of geometry and physics and one can find, in the literature,
considerations giving reasons for the four dimensionality of space-time
which are related to effects calculated for some physical law (for a review,
see [52]). Besides, various heuristic reasons may be given for space-time to
have this dimension. However, in spite of all this, the mechanism (if it exists)
responsible for fixing this effective number of dimensions  to four is still a
mystery. As a matter of fact, in some physical theories, other dimensions have
been considered. In certain cases, they are used as a purely mathematical
trick (see, for example,  [53]). In other cases, the fact
 that the dimension of space-time differs from four has both a practical and a
theoretical interest and is taken to descibe physical reality as it is in models
based on the original suggestion of Kaluza and Klein with the aim of unifying
all known interactions (see, for example, [54]), or in the framework of the
physics of low dimensions quantum structures  [55].
This new branch of physics,
which has proved as important for fundamental research as for the applied
one, has numerous theoretical links with other domains of physics and
maintains  close relations between experimental  works and advanced
technology such as the domain of numerous new materials produced and
studied nowadays among which we can mention magnetic materials and
electronic  components. As for the space sub-jacent to the internal
structure of matter, there is nothing to prevent it from having a given
 dimension, especially that it has been introduced, from the beginning
(i.e from the introduction of the isospin [9]), essentially by analogy with the
space-time. Moreover, to our knowledge, no experiment mentions any observable
difference between the relativistic and non-relativistic intrinsic internal
structures. In this context, it would be interesting to think of other dimensions
for this structure, but always in conformity with the New-Stein symmetry. In
this perspective, we are encouraged by the fact that several two and three-dimensional space-time systems have been intensely studied over the last
two decades, especially when the structure of relevant phenomena is
effectively confined in two or one spatial dimensions. Some of these models
have already led to immediate and spectacular applications. Among these
phenomena   and models, we may quote [55], [56] : non-linear optics, theory of
anyons and its connection with the fractional quantum Hall effect, Chern-Simmons gauge theories,
high  temperature superconductivity... Besides, according
to Salam [16],  the Theory of Strings must rather be considered as equivalent
to a theory of fields in a two-dimensional space-time. All this has, in fact,
recently led to the study of various two or three-dimensional space-time
symmetry groups [57] such as : Galilei,  Galilei-Similitude, Schr\"odinger,
 Poincar\'e and Conformal groups.

In this perspective, let us consider the  New-Stein group $G_2$ associated with
a  two-dimensional internal space. This is achieved by  replacing, in the definition
of ${\cal G, H}_3$ (resp ${\cal SU}(2)$) by ${\cal H}_2$ (resp ${\cal SO}(2))$.  This dimension change
will induce a radical change in interpretation and in structure. In fact, the isospin
in two dimensions differs fundamentally from isospin in higher dimensions,
because the angular momentum algebra in two dimensions is the trivial
commutative algebra $\R$ ;   there is no analog  of the quantization of
angular momentum in higher dimensions, associated with the ${\cal S O}(n)$
rotation algebra, $n \geq 3$. Thus, in the $G_2$ context, a particle can have arbitrary
real isospin, like the anyon which may be considered as being a particle
with arbitrary spin and statistics [56].
 It would be interesting to explore , in the
framework of low dimensions internal structure, the new ideas linked with
the anyons physics and, especially, those associated with phase transitions
and  ``critical phenomena'' of small-dimensional systems [55], [58]. As for the
structural change, it lies in the fact that $G_2$ has significantly richer extensions
than $G$,  since the space of  equivalence classes  of the central extensions of
${\cal G}_2$ (resp $\cal G$) by $\R\;$ is of dimension thirteen (resp. eleven). In a
near perspective, it appears that the various  symmetries associated with the bidimensional
internal space differ fundamentally from those associated with the
other dimensions, since the equivalence classes space  of the infinitesimal
exponents of $F_n (n = 1$ or $n \geq 3)$ is, as that of the Einstein group [26],
 unidimensional, whereas that of
$F_2$ is of dimension four. Consequently, it would be interesting to
include the other internal symmetries, defined by the other central extensions
 of $F_2$,  in a fundamental global symmetry, in similitude with that studied in
this paper, which we can call a New-Stein-type symmetry.
\vspace{3mm}\\
{\bf Dedication}
\vspace{2mm}

We dedicate this paper to the spirit of Mosh\'e Flato, teacher and friend, in
gratitude for many stimulating conversations regarding mass problem and for
 his careful reading of  the French version of the manuscript some months
 before his sudden disappearance.
\vspace{3mm}\\
{\bf References}
\vspace{2mm}

{ \footnotesize 1.
Glashow, S.L., {\it Nucl. Phys}. {\bf 22}, 579 (1961) ; Salam,A., in  W. Svartholm
(ed.), {\it Proceedings of the Eight Nobel Symposium}, Almquist and Wiskell,
Stockholm, 1968, p. 367 ; Weinberg, S., {\it Phys. Rev. Lett}. {\bf 19},
1264 (1967).

2.
Collins, P.D.B., Martin,A.D., Squires,E.J., {\it Particle Physics and Cosmology},
John Wiley and Sons, New York, 1989.

3.
Kolb, E.W., Turner, M.S., {\it The Early Universe}, Addison-Wesley Publ. Co.,
Reading, Mass., 1990.

4.
Feynman, R., {\it QED,  the Strange Theory of Light and Matter,} Princeton
University Press, Princeton, N.J., 1985.

5.
Nambu,Y., Jonalasinio,G., {\it Phys. Rev.} {\bf 122}, 345 (1961) ; Higgs, P. W.,
{\it Phys. Rev.} {\bf 145}, 1156 (1966).

6.
Pauli, W.,Touschek, B., {\it Nuovo Cimento} {\bf 14}, 205 (1959).

7.
Heisenberg,W., in J. Mehra (ed.),{\it The Physicist's Conception of Nature},
D. Reidel Publishing Company, Dordrecht, Holland, 1973, p. 264.

8.
Heisenberg,W., {\it Naturwiss.} {\bf 63}, 1 (1976).

9.
Heisenberg,W., {\it Z. Physik} {\bf 77}, 1 (1932) ; {\bf 78}, 156 (1932).

10.
Enz,C.P., {\it W. Pauli's Scientific Work}, in J. Mehra (ed.), {\it The Physicist's
Conception of Nature}, D. Reidel Publishing Company, Dordrecht, Holland, 1973,
 p. 766.

11.
D\"urr, H.P., Heisenberg, W.,Mitter, H.,Schlieder, S.,Yamazaki,K.,
{\it Z. Naturforsch.} {\bf 14a}, 441 (1959) ; Heisenberg, W., {\it Introduction to
the Unified Field Theory of Elementary Particles}, Interscience Publishers, London,
1966.

12.
Campagnari,C., Franklin,M., {\it Rev. Mod. Phys.} {\bf 69}, 137 (1997).

13.
Pati,J.C., Salam, A.,Strathdee, J., {\it Nucl. Physics} {\bf B185}, 445 (1981).

14.
Langacker,P., {\it Phys. Reports} {\bf 72}, 185 (1981).

15.
Weinberg,S., {\it Phys. Rev. Lett.} {\bf 42}, 850 (1972).

16.
Dirac,P.A.M., {\it Methods in Theoretical  Physics}, Lecture given in 1968 at
International Centre for Theoretical Physics, Trieste and reprinted in Salam,
A., {\it Unification of Fundamental Forces}, Cambridge University Press,
Cambridge, 1990, p. 125.

17.
Linde,A.D., {\it Particle Physics and Inflationnary Cosmology}, Harwood
Academic Publishers, London, 1990.

18. Guth,A.H., {\it Phys. Rev.} {\bf D23}, 347 (1981) ; for a
recent outlook see : Turner, M.S., astro-ph/0212281 and references
therein.

19.
Quoted in Gamow,G., {\it My World Line : An Informal Autobiography}, Viking
Press, New York, 1970, where Gamow reports the astonishment of Einstein
when learning about Jordan's hypothesis.

20.
Tryon,E.P., {\it Nature} {\bf 246}, 396 (1973).

21.
Moshinsky,M., {\it Harmonic Oscillators in Modern Physics}, Gordon and Breach,
New York, 1969.

22.
Kim,Y.S.,Noz,M.E., {\it Theory and Applications of the Poincar\'e Group},
D. Reidel Publishing  Company, Dordrecht, Holland, 1986.

23.
Yukawa,H, {\it Phys. Rev.} {\bf 91}, 416 (1953) ; Feynman, R.P., Kislinger,M.,
Ravndal, F., {\it Phys. Rev.} {\bf D3}, 2706 (1971) ; Takabayasi,T.,
{\it Prog. Theor. Phys. Suppl.} {\bf 67}, 1 (1979).

24.
Born,M., {\it Nature} {\bf 163}, 207 (1949); {\it Rev. Mod. Phys.} {\bf 21},
 463 (1949),  see also {\it Albert Einstein / Max Born Briefwechsel 1916-1955},
Nymphenburger Verlagshandlung GmbH., Munich, 1969.

25.
Beau,D., Horchani,S., {\it J. Math. Phys.} {\bf 20}, 1700 (1979).

26.
Horchani,S., {\it Ann. Inst. Henri Poincar\'e} {\bf 15}, 321 (1971).

27.
Horwitz,L.P.,Piron, C.,Reuse, F., {\it Helv. Phys. Acta} {\bf 48}, 546 (1975) ;
Giovannini, N., Piron, C., {\it Helv. Phys. Acta} {\bf 52}, 518 (1979).

28.
Charfi,A., {\it Sur la Cohomologie et les Repr\'esentations Non Lin\'eaires des
Groupes de Weyl et des Groupes de Newton}, Th\`ese de Doctorat d'\'Etat, D\'epartement
de Math\'ematiques de la Facult\'e des Sciences de l'Universit\'e de Tunis II,
 Avril 1992.

29.
Fermi,E. {\it Nuclear Physics}, Chicago University Press, Chicago, 1949.

30.
Yang,C. N., in {\it Nobel Lectures, Physics, 1942-1962}, Elsevier
Publishing Company, Amsterdam, 1964, p. 393.

31.
Beau,D., {\it J. Math. Phys.} {\bf 24}, 1299 (1983).

32.
Horwitz,L.P.,Piron,C., {\it Helv. Phys. Acta} {\bf 46}, 316 (1973).

33.
D'Souza,A., Kalman,C.S., {\it Preons, Models of Leptons, Quarks and Gauge
Bosons as Composite Objects}, World Scientific, Singapore, 1992.

34.
For a review of these models see, for example, Hinchliffe, I.,
{\it Ann. Rev. Nucl. and Part. Sci.} {\bf 36}, 505 (1986).

35.
de Broglie, L., {\it Une nouvelle conception de la lumi\`ere}, Hermann,
Paris, 1934 ; {\it Une nouvelle th\'eorie de la lumi\`ere : La m\'ecanique ondulatoire
du photon}, 2 volumes, Hermann, Paris, 1941, 1942.

36.
Flato,M.,Fronsdal,C., {\it Phys. Lett.} {\bf 97B}, 236 (1980).

37.
Flato, M, Fronsdal,C., {\it J. Math. Phys.} {\bf 32}, 524 (1991).

38.
Mackey,G.W., {\it The Theory of Unitary Group Representations},  The University
of Chicago Press, Chicago, 1976.

39.
Nachbin,L., {\it The Haar Integral}, Van Nostrand, New York, 1965.

40.
For a review and a bibliography on the deformations theory,  see,  for example,
Conatser,C.W., in V. Komkov (ed.), {\it Texas Tech. University, Mathematics
Series $N^o$ 10},  Lubbock, Texas, 1972, p. 107.

41.
Horchani,S., {\it C. R. Acad. Sc. Paris} {\bf 277}, 201 (1973).

42.
Chevalley,C., Eilenberg,S., {\it Trans. Am. Math. Soc.} {\bf 63}, 85 (1948) ;
Hochschild,G., Serre, J.P., {\it Ann. of Math.} {\bf 57}, 591 (1953).

43.
Bourbaki,N., {\it Groupes et Alg\`ebres de Lie, Chapitre 1},  Hermann, Paris, 1960.

44.
Jacobson,N., {\it Lie Algebras}, John Wiley and Sons, New York, 1966.

45.
Stueckelberg,E.C.G., {\it Helv. Phys. Acta} {\bf 14}, 588 (1941) ;
Feynman, R.P., {\it Phys. Rev.} {\bf 76}, 749 (1949) ; {\bf 80}, 440 (1950) ;
Nambu, Y., {\it Prog. Theor. Phys.} {\bf 5}, 82 (1950) ; Schwinger, J.,
{\it Phys. Rev.} {\bf 82}, 664 (1951).

46.
Weinberg,S., in Feynman,R., Weinberg,S., {\it Elementary Particles and the Laws
 of Physics}, Cambridge University Press, Cambridge, 1987, p. 67.

47. In honor of the philosopher Ibrahim ibn Saiyar  Al-Nazzam
(Bassorah 775- Baghdad 846) who, in his treatise on motion [{\it
Kitab fi alharaka}],  tried to solve Zeno's paradox by asserting
that the apparently continuous motion of macroscopic bodies is, in
reality, the
 combination of a ``leap'' [{\it tafrah}] microscopic motions sequence constituting
``an unanalyzable interphenomenon''. This led Max Jammer to assert, in
48.  p. 259,  that ``Al-Nazzam's notion of leap, his designation of an unanalyzable
 interphenomenon, may be regarded as an early forerunner of Bohr's conception
of quantum jumps''.

48. Jammer,M., {\it The Philosophy of Quantum Mechanics, The
Interpretations of Quantum Mechanics in Historical Perspective},
John Wiley and Sons, New York, 1974.

49.
 Green, M.B., Schwarz, J.H.,Witten, E., {\it Superstring
Theory}, Cambridge University Press, Cambridge, 1987.

50.
Weinberg,S., {\it Phys. Rev.} {\bf D9 }, 3357 (1974).

51.
Linde,A.D., {\it Rep. Prog. Phys.} {\bf 42}, 389 (1979).

52.
Barrow,J.O., {\it Philos. Trans. R. Soc. London, Ser.A}{\bf 310}, 337
(1983).

53.
t'Hooft,G., Veltmann,M., {\it Nucl. Phys.} {\bf B44}, 189 (1972).

54.
Appelquist,T.,Chodos,A., Freund,P.G.O., (eds), {\it Modern Kaluza-Klein
Theories}, Frontiers in Physics, Addison-Wesley, Reading, MA, 1987.

55.
Duff,M. J., Pope,C.N., Sezgin,E., (eds), {\it Proceedings of the  Trieste Conference
on Supermembranes and Physics in 2+1 Dimensions, July 17-21, 1989, ICTP},
World Scientific, Singapore, 1990 ; Thouless, D., in {\it The New Physics},
P.Davies (ed.), Cambridge
 University Press, Cambridge , 1993, p. 209 ; Stormer, H.L.,
{\it Rev. Mod. Phys.}  {\bf 71}, 875 (1999).

56.
Moore,G.,Seiberg, N., {\it Phys. Lett.} {\bf 220B}, 422 (1989) ;
Fr\"ohlich,J.,Marchetti,P.A., {\it Commun. Math. Phys.} {\bf 121}, 177 (1989) ;
Jackiw, R.,Nair,V.P., {\it Phys. Rev.} {\bf D43}, 193 (1991).

57.
Rideau,G.,Winternitz,P., {\it J. Math. Phys.} {\bf 31}, 1096 (1990);
{\bf 34},  558 (1993) ; Grigore,D.R., {\it J. Math. Phys.} {\bf 34}, 4172 (1993) ;
Bose,S.K., {\it Commun. Math. Phys.} {\bf 169}, 385 (1995).

58.
Barber,M.N., {\it Phys. Rep.} {\bf 59} , 375 (1980).}
\end{document}